\documentclass[10pt]{iopart}

\usepackage{graphics}
\usepackage{graphicx}
\usepackage{multirow}
\usepackage{bm}
\usepackage{harvard}
\begin{document}

\title{A Multimodal Approach to Estimating Vigilance Using EEG and Forehead EOG}

\author{Wei-Long Zheng$^{1}$ and Bao-Liang Lu$^{1,2,3,\dag}$}

\address{$^1$Center for Brain-like Computing and Machine Intelligence, Department of Computer Science and Engineering, Shanghai Jiao Tong University, Shanghai, China}
\address{$^2$Key Laboratory of Shanghai Education Commission for Intelligent Interaction and Cognitive Engineering, Shanghai Jiao Tong University, Shanghai, China}
\address{$^3$Brain Science and Technology Research Center, Shanghai Jiao Tong University, Shanghai, China}
\address{$^\dag$Author to whom any correspondence should be addressed.}
\ead{weilong@sjtu.edu.cn, bllu@sjtu.edu.cn}
\vspace{10pt}
\begin{indented}
\item[]August 2016
\end{indented}

\begin{abstract}
\newline
\textit{Objective.} Covert aspects of ongoing user mental states
provide key context information for user-aware human computer
interactions. In this paper, we focus on the problem of estimating the
vigilance of users using EEG and EOG signals. \textit{Approach.} To
improve the feasibility and wearability of vigilance estimation
devices for real-world applications, we adopt a novel electrode
placement for forehead EOG and extract various eye movement features,
which contain the principal information of traditional EOG. We explore
the effects of EEG from different brain areas and combine EEG and
forehead EOG to leverage their complementary characteristics for
vigilance estimation. Considering that the vigilance of users is a
dynamic changing process because the intrinsic mental states of users
involve temporal evolution, we introduce continuous conditional neural
field and continuous conditional random field models to capture
dynamic temporal dependency. \textit{Main results.} We propose a
multimodal approach to estimating vigilance by combining EEG and
forehead EOG and incorporating the temporal dependency of vigilance
into model training. The experimental results demonstrate that
modality fusion can improve the performance compared with a single
modality, EOG and EEG contain complementary information for vigilance
estimation, and the temporal dependency-based models can enhance the
performance of vigilance estimation. From the experimental results, we
observe that theta and alpha frequency activities are increased, while
gamma frequency activities are decreased in drowsy states in contrast
to awake states. \textit{Significance.} The forehead setup allows for the
simultaneous collection of EEG and EOG and achieves comparative performance using only four shared electrodes in comparison with the temporal and posterior sites.
\end{abstract}

%
%
\submitto{\JNE}
%
\maketitle
%
\ioptwocol
\section{Introduction}

Humans interact with their surrounding complex environments based on
their current states, and context awareness plays an important role
during such interactions. However, the majority of the existing
systems lack this ability and generally interact with users in a
rule-based fashion. Covert aspects of ongoing user mental states
provide key context information in user-aware human computer
interactions \cite{zander2012context}, which can help systems react
adaptively in a proper manner. Various studies have introduced the
assessment of the mental states of users, such as intention, emotion,
and workload, to promote active interactions between users and
machines
\cite{muhl2014eeg,lu2015combining,kang2015human,zheng2015investigating}. Zander
and Kothe proposed the concept of a passive brain-computer interface
(BCI) to fuse conventional BCI systems with cognitive monitoring
\cite{zander2011towards}. It is attractive to implement these novel
BCI systems with increasing information flow of human states without
simultaneously increasing the cost significantly. Among these cognitive states, vigilance is a vital component, which refers to the ability to endogenously maintain focus.

Various working environments require sustained high vigilance,
particularly for some dangerous occupations such as driving trucks and
high-speed trains. In these cases, a decrease in vigilance
\cite{grier2003vigilance} or a momentary lapse of attention \cite{peiris2011detection,davidson2007eeg} might severely endanger public transportation safety. Driving fatigue is reported to be a major factor in fatal road accidents.

Various approaches for estimating vigilance levels have been proposed
in the literature
\cite{ji2004real,dong2011driver,sahayadhas2012detecting}. However,
several research challenges still exist. Vigilance decrement is a
dynamic changing process because the intrinsic mental states of users
involve temporal evolution rather than a time point. This process
cannot simply be treated as a function of the duration of time while
engaged in tasks. The ability to predict vigilance levels with high
temporal resolution is more feasible in real-world applications
\cite{davidson2007eeg}. Moreover, drivers' vigilance levels cannot be
simply classified into several discrete categories but should be
quantified in the same way as the blood alcohol level
\cite{dong2011driver,ranney2008driver}. We still lack a standardized
method for measuring the overall vigilance levels of humans.

Among various modalities, EEG is reported to be a promising
neurophysiological indicator of the transition between wakefulness and
sleep in various studies because EEG signals directly reflect human
brain activity
\cite{berka2007eeg,khushaba2011driver,shi2013eeg,lin2014wireless,martel2014eeg,kim2014detection}. Rosenberg
and colleagues recently presented a neuromarker for sustained
attention from whole-brain functional connectivity
\cite{rosenberg2015neuromarker}. They developed a network model called
the sustained attention network for predicting attentional
performance. Moreover, EEG has intrinsic potential to allow fatigue
detection at onset or even before onset
\cite{davidson2007eeg}. O'Connell and colleagues examined the temporal
dynamics of EEG signals preceding a lapse of sustained attention
\cite{o2009uncovering}. Their results demonstrated that the specific
neural signatures of attentional lapses are registered in the EEG up
to 20 s prior to an error. Lin \textit{et al.} presented a wireless and wearable EEG system for evaluating drivers' vigilance levels, and they tested their system in a virtual driving environment \cite{lin2014wireless}. They also combined lapse detection and feedback efficacy assessment for implementing a closed-loop system. By monitoring the changes of EEG patterns, they were able to detect driving performance and estimate the efficacy of arousing warning feedback delivered to drowsy subjects \cite{lin2013can}.

\begin{figure*}[!htbp]
 \centering
 \includegraphics[width=1.88\columnwidth]{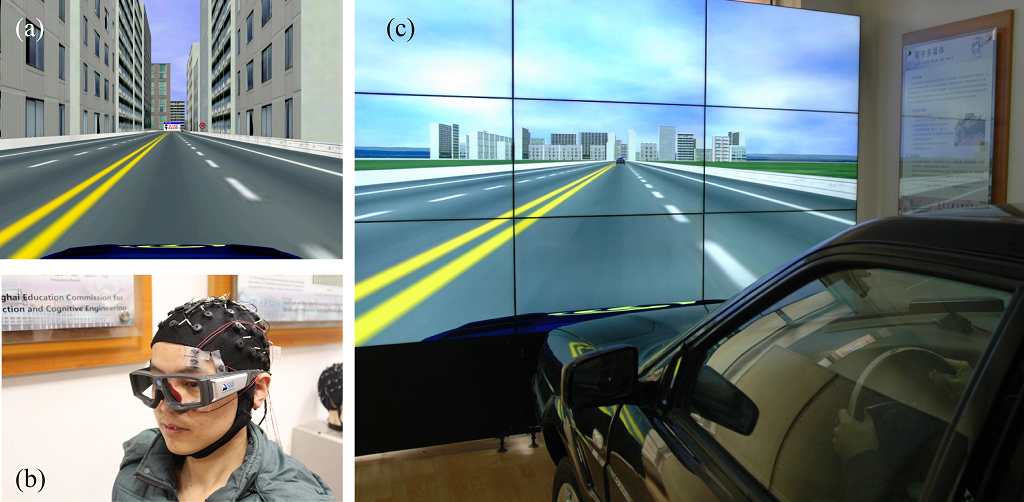}
 \caption{The simulated driving system and the experimental scene. (a)
   The virtual-reality-based simulated driving scenes, including
   various weather and roads. (b) Forehead EOG, EEG and eye movements
   are simultaneously recorded using the Neuroscan system and eye
   tracking glasses. (c) The simulated driving experiments are performed in a real vehicle without unnecessary engine and other components. During the experiments, the subjects are asked to drive the car using the steering wheel and gas pedal. The driving scenes are synchronously updated according to subjects' operations. There is no warning feedback to subjects after sleeping.}
\label{fig:figure4}
\end{figure*}

In addition to EEG, EOG signals contain characteristic information on
various eye movements, which are often utilized to estimate vigilance
because of its easy setup and high signal-noise ratio \cite{papadelis2007monitoring,damousis2008fuzzy,ma2010vigilance,ma2014eog}. Researchers have developed various multimodal approaches for constructing hybrid BCIs \cite{pfurtscheller2010hybrid} and combining brain signals and eye movements for robotic control and cognitive monitoring \cite{lee2010brain,mcmullen2014demonstration,zheng2014multimodal,ma2014eog,lu2015combining}. Simola \textit{et al.} studied the valence and arousal interactions under free viewing of emotional scenes by analysing eye movement behaviours and eye-fixation-related potentials \cite{simola2015affective}. Their findings support the multi-dimensional, interactive model of emotional processing. Moreover, Bulling and colleagues found that eye movements from EOG signals are good indicators for activity recognition \cite{bulling2011eye}. However, the electrodes in the traditional EOG are placed around the eyes, which may distract users and cause discomfort. In our previous study, we proposed a new electrode placement on the forehead and extracted various eye movement features from the forehead EOG \cite{zhang2015novel,huo2016driving}. Various studies have indicated that signals from different modalities represent different aspects of convert mental states \cite{calvo2010affect,sahayadhas2012detecting,d2015review}. EEG and EOG represent internal cognitive states and external subconscious behaviours, respectively. These two modalities contain complementary information and can be integrated to construct a more robust vigilance estimation model.

In this paper, we present a multimodal approach for vigilance
estimation by combining EEG and forehead EOG. The main contributions
of this paper are as follows: 1) we explore the effect of EEG for
vigilance estimation in different brain areas: frontal, temporal, and
posterior; 2) we propose a multimodal vigilance estimation framework
with EEG and forehead EOG in terms of feasibility and accuracy; 3) we
acquire both EEG and EOG signals simultaneously with four shared
electrodes on the forehead and combine them for vigilance estimation;
4) we reveal the complementary characteristics of EEG and forehead EOG
modalities for vigilance estimation; 5) we apply continuous
conditional neural field (CCNF) and continuous conditional random
field (CCRF) models to enhance the performance of the vigilance estimation model to capture dynamic temporal dependency; and 6) we investigate neural patterns regarding critical frequency activities under awake and drowsy states.

\section{Methods}
\subsection{Experiment Setup}

To collect EEG and EOG data, we developed a virtual-reality-based
simulated driving system. A four-lane highway scene is shown on a
large LCD screen in front of a real vehicle without the unnecessary
engine and other components. The vehicle movements in the software are
controlled by the steering wheel and gas pedal, and the scenes are
simultaneously updated according to the participants' operations. The
road is primarily straight and monotonous to induce fatigue in the subjects more easily. The simulated driving system and the experimental scene are shown in Figure \ref{fig:figure4}.

A total of 23 subjects (mean age: 23.3, STD: 1.4, 12 females)
participated in the experiments. All participants possessed normal or
corrected-to-normal vision. Caffeine, tobacco, and alcohol were
prohibited prior to participating in the experiments. At the beginning of the experiments, a short pre-test was performed to ensure that every participant understood the instructions. Most experiments were performed in the early afternoon (approximately 13:30) after lunch to induce fatigue easily when the circadian rhythm of sleepiness reached its peak \cite{ferrara2001much}. The duration of the entire experiment was approximately 2 hours. The participants were asked to drive the car in the simulated environments without any alertness.

\begin{figure}[!htbp]
 \centering
 \includegraphics[width=\columnwidth]{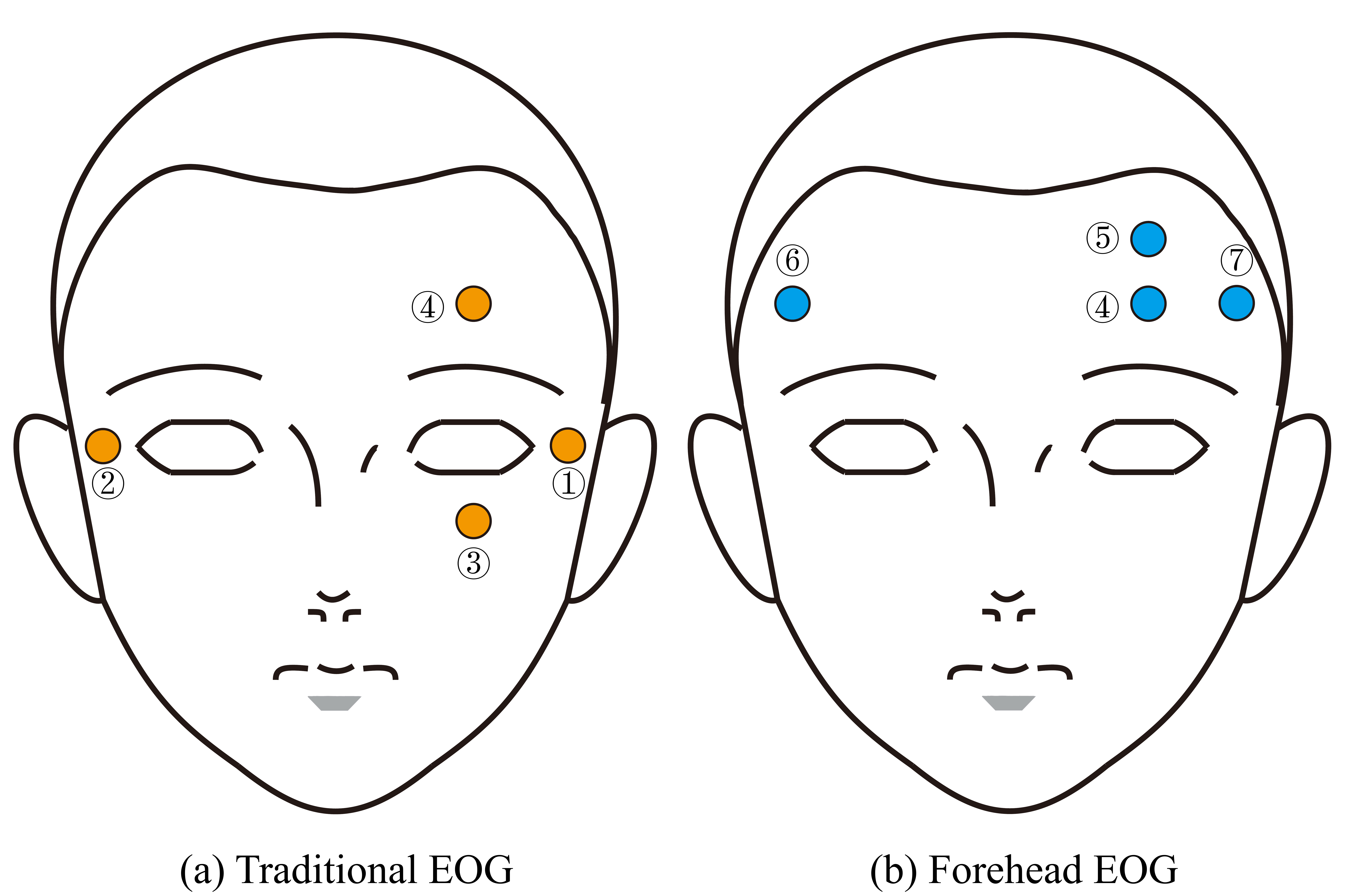}
 \caption{Electrode placements for the traditional and forehead EOG setups. The yellow and blue dots indicate the electrode placements of the traditional EOG and forehead EOG, respectively. Electrode four is the shared electrode of both setups.}
\label{fig:figure1}
\end{figure}

\begin{figure}[!htbp]
 \centering
 \includegraphics[width=0.75\columnwidth]{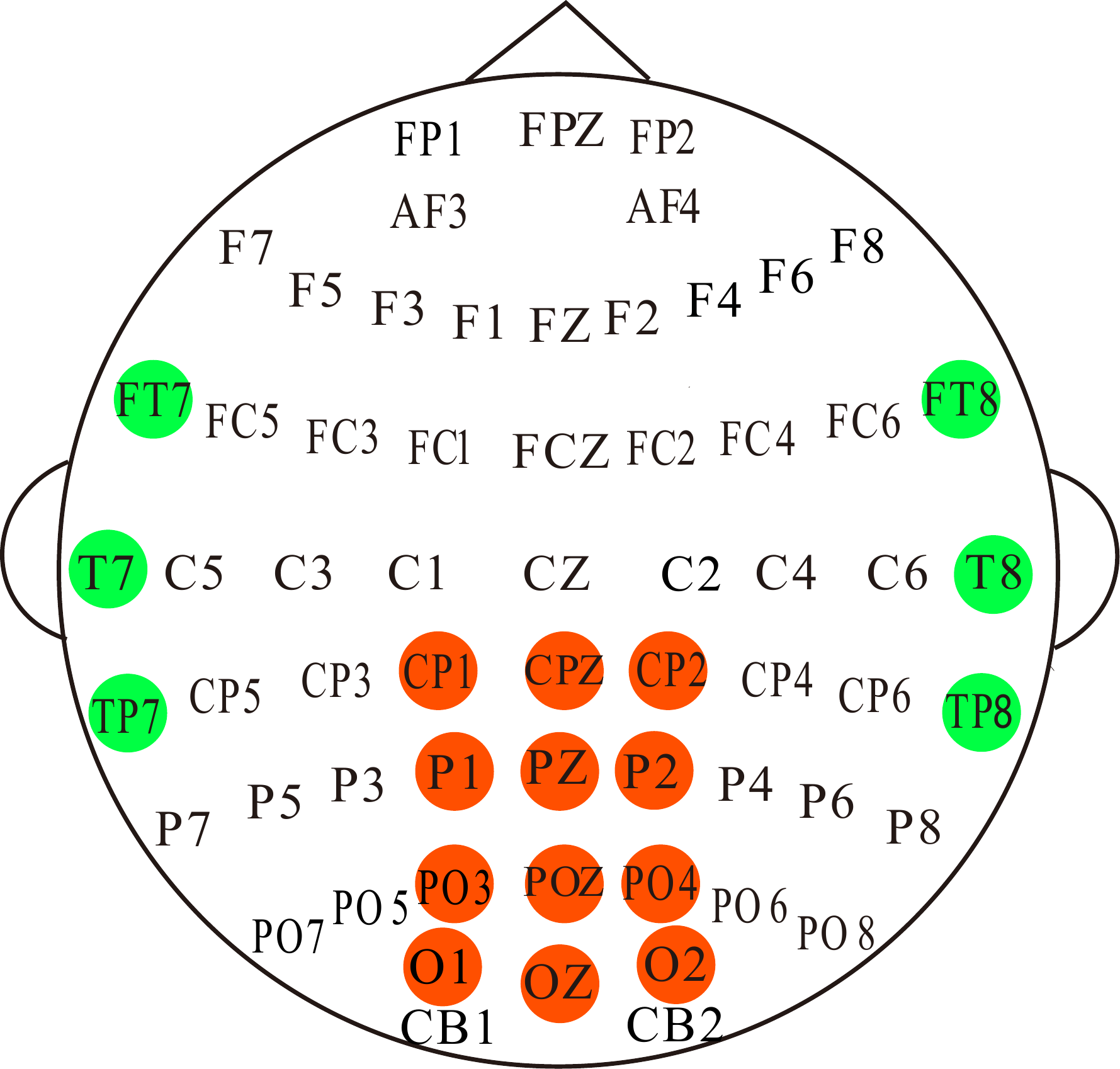}
 \caption{Electrode placements for the EEG setups. 12-channel and 6-channel EEG signals were recorded from the posterior site (red colour) and temporal site (green colour), respectively.}
\label{fig:electrode}
\end{figure}

Both EEG and forehead EOG signals were recorded simultaneously using
the Neuroscan system with a 1000 Hz sampling rate. The electrode
placement of the forehead EOG \cite{zhang2015novel} is shown in Figure
\ref{fig:figure1}. For the EEG setup, we recorded 12-channel EEG
signals from the posterior site ($CP1$, $CPZ$, $CP2$, $P1$, $PZ$,
$P2$, $PO3$, $POZ$, $PO4$, $O1$, $OZ$, and $O2$) and 6-channel EEG
signals from the temporal site ($FT7$, $FT8$, $T7$, $T8$, $TP7$, and
$TP8$) according to the international 10-20 electrode system shown in
Figure \ref{fig:electrode}. Eye movements were simultaneously recorded using SMI ETG eye tracking glasses\footnote{\url{http://eyetracking-glasses.com/}}, and the facial video was recorded from a video camera mounted in front of the participants.

For reproducing the results in this paper and enhancing cooperation in related research fields, the dataset used in this study will be freely available to the academic community as a subset of SEED\footnote{\url{http://bcmi.sjtu.edu.cn/~seed/}}.

\subsection{Vigilance Annotations}
The primary challenge of vigilance estimation using a supervised
machine learning paradigm is how to quantitatively label the sensor
data because the ground truth of convert mental states cannot be
accurately obtained in theory. To date, researchers have proposed
various vigilance annotation methods in the literature, such as lane
departure and local error rates
\cite{wang2015eeg,makeig1993lapse}. Lin \textit{et al.} designed an
event-related lane-departure driving task in which the subjects were
asked to respond to the random drifts as soon as possible and the
response time reflected the vigilance states of the subjects
\cite{lin2010tonic,lin2013can}. Shi and Lu \cite{shi2013eeg} conducted
a study in which the local error rate of the subjects' performance was
used as the vigilance measurement. The subjects were asked to press correct buttons according to the colours of traffic signs. These two annotation methods are based on subjects' behaviours and can reflect their actual vigilance levels to some extent. However, they are not feasible for dual tasks, particularly in real-world driving environments.

There is another annotation method called PERCLOS
\cite{dinges1998perclos}, which refers to the percentage of eye
closure. It is one of the most widely accepted vigilance indices in
the literature
\cite{trutschel2011perclos,bergasa2006real,dong2011driver}. Conventional
driving fatigue detection methods utilize facial videos to calculate
the PERCLOS index. However, the performance of facial videos can be
influenced by environmental changes, especially for various
illuminations and heavy occlusion. In this study, we adopt an
automatic continuous vigilance annotation method using eye tracking
glasses, which was proposed in our previous work
\cite{gao2015evaluating}. This approach allows  vigilance to be
measured in both laboratory and real-world environments.

\begin{figure}[!htbp]
 \centering
 \includegraphics[width=\columnwidth]{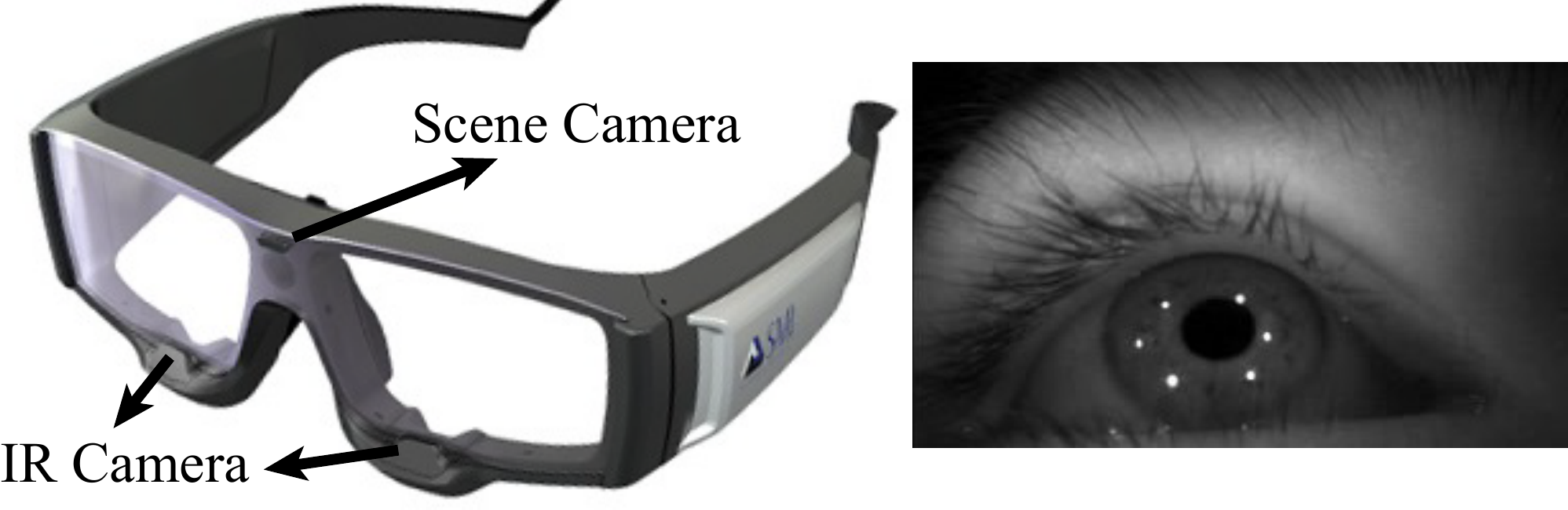}
 \caption{The SMI eye tracking glasses used in this study and the pupillary image captured in one experiment.}
\label{fig:etg}
\end{figure}

Compared with facial videos, eye tracking glasses can more precisely capture different eye movements, such as blink, fixation, and saccade, as shown in Figure \ref{fig:etg}. The eye tracking-based PERCLOS index can be calculated from the percentage of the durations of blinks and `CLOS' over a specified time interval as follows:
\begin{equation}
PERCLOS = \frac{blink+CLOS}{interval}, and
\end{equation}
\begin{equation}
interval = blink+fixation+saccade+CLOS,
\end{equation}
where `CLOS' denotes the duration of the eye closures.

We evaluated the efficiency of the eye tracking-based method for vigilance annotations with the facial videos recorded simultaneously and found a high correlation between the PERCLOS index and the subject's current cognitive states. Compared with other approaches \cite{shi2010off,ma2014eog,wang2015eeg}, this method is more feasible for real-world driving environments, where performing dual tasks can distract attention and cause safety issues \cite{oken2007sleeping}. This new vigilance annotation method can be performed automatically without too much interference to the drivers.

Note that although the eye tracking-based approach can estimate the
vigilance level more precisely, it is not currently feasible to apply it to real-world applications due to its very expensive cost. Here, we utilize eye tracking glasses as a vigilance annotation device to obtain more accurate labelled EEG and EOG data for training vigilance estimation models.

\subsection{Feature Extraction}
\subsubsection{Preprocessing for Forehead EOG}
For traditional EOG recordings, the electrodes are mounted around the
eyes using the electrodes numbered one to four in Figure
\ref{fig:figure1} (a). However, in real-world applications, such
electrode placement is not easily mounted and may distract users with
discomfort. To implement wearable devices for real-world vigilance
estimation, we propose placing all the electrodes on the forehead, as
shown in Figure \ref{fig:figure1} (b), and separating vertical EOG
(VEO) and horizontal EOG (HEO) using the electrodes numbered four to
seven shown in Figure \ref{fig:figure1} (b). For the traditional EOG setup shown in Figure \ref{fig:figure1} (a), the VEO and HEO signals are obtained by subtracting electrodes four and three and electrodes one and two, respectively. VEO and HEO signals contain details of eye movements, such as blink, saccade, and fixation.

How to extract VEO and HEO signals from the forehead EOG setup is one of the key problems in this study. We extracted VEO$_{\mbox{\scriptsize \emph{f}}}$ signals from electrodes numbered four and seven and extracted HEO$_{\mbox{\scriptsize \emph{f}}}$ signals from electrodes five and six using two separation strategies: the minus rule and independent component analysis (ICA). For the minus rule, the subtraction of channels five and seven is an approximation of VEO, named VEO$_{\mbox{\scriptsize \emph{f}}}$, and the subtraction of channels five and six is an approximation of HEO, named HEO$_{\mbox{\scriptsize \emph{f}}}$. Here, the subscript `$f$' indicates `forehead'. 

\begin{figure}[!b]
 \centering
 \includegraphics[width=\columnwidth]{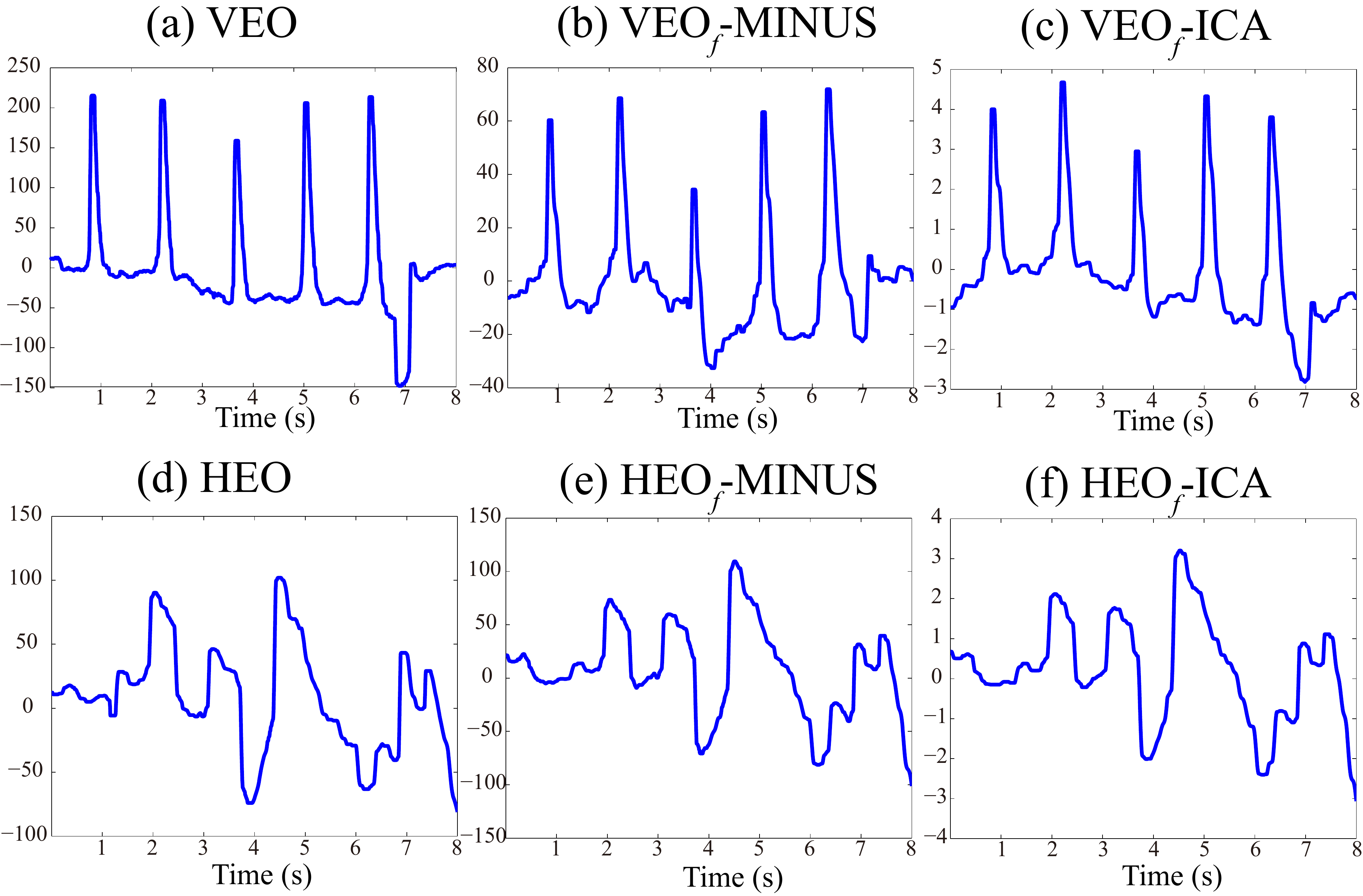}
 \caption{Comparison of traditional EOG and forehead EOG using minus
   operation and ICA separation strategies. Here, (a) and (d) are
   traditional VEO and HEO; (b) and (e) are extracted
   VEO$_{\mbox{\scriptsize \emph{f}}}$ and HEO$_{\mbox{\scriptsize
       \emph{f}}}$ from forehead EOG using the minus operation; and (c)
   and (f) are extracted VEO$_{\mbox{\scriptsize \emph{f}}}$ and
   HEO$_{\mbox{\scriptsize \emph{f}}}$ from forehead EOG using the ICA approach.}
\label{fig:figure2}
\end{figure}

ICA is a blind source separation method proposed to decompose a
multivariate signal into independent non-Gaussian signals
\cite{delorme2004eeglab}. We extracted the VEO$_{\mbox{\scriptsize
    \emph{f}}}$ and HEO$_{\mbox{\scriptsize \emph{f}}}$ components
using FASTICA \cite{delorme2004eeglab} from channels four and seven
and channels five and six, respectively. The comparison of the
traditional EOG and forehead EOG using the minus operation and ICA
separation strategies is depicted in Figure \ref{fig:figure2}. As shown, the extracted VEO$_{\mbox{\scriptsize \emph{f}}}$ and HEO$_{\mbox{\scriptsize \emph{f}}}$ from the forehead electrodes have similar waves to the traditional ones, and the forehead VEO$_{\mbox{\scriptsize \emph{f}}}$ and HEO$_{\mbox{\scriptsize \emph{f}}}$ can capture critical eye movements, such as blinks and saccades.

\begin{figure}[!t]
 \centering
 \includegraphics[width=\columnwidth]{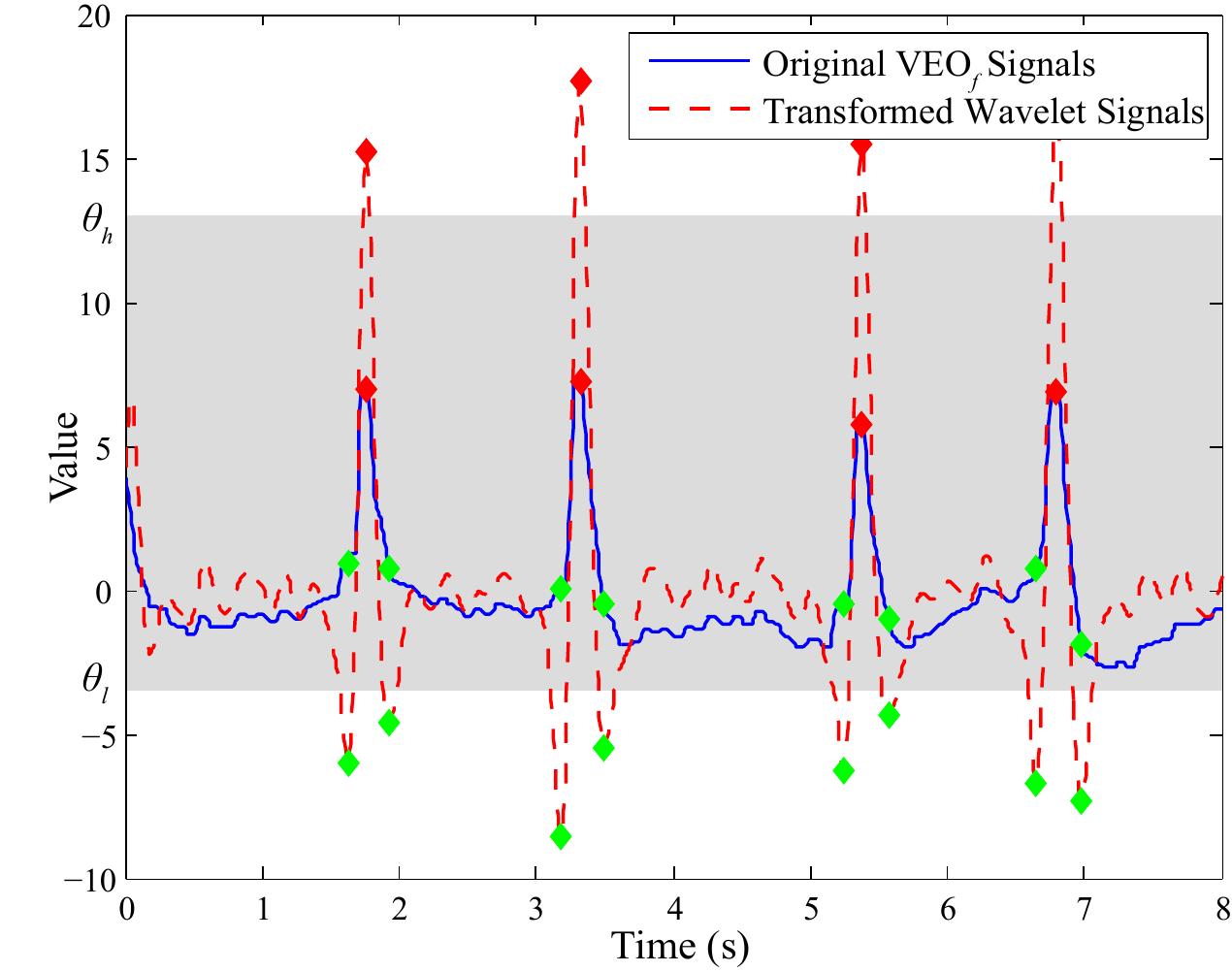}
 \caption{The blink detected by using continuous wavelet transform. We applied two thresholds $\theta_{h}$ and $\theta_{l}$ on the transformed wavelet signals and detected peaks to locate blink segments. Red markers indicate the peaks of each blink, and green markers indicate the start and end points of each blink.}
\label{fig:blink}
\end{figure}

\begin{figure}[!t]
 \centering
 \includegraphics[width=\columnwidth]{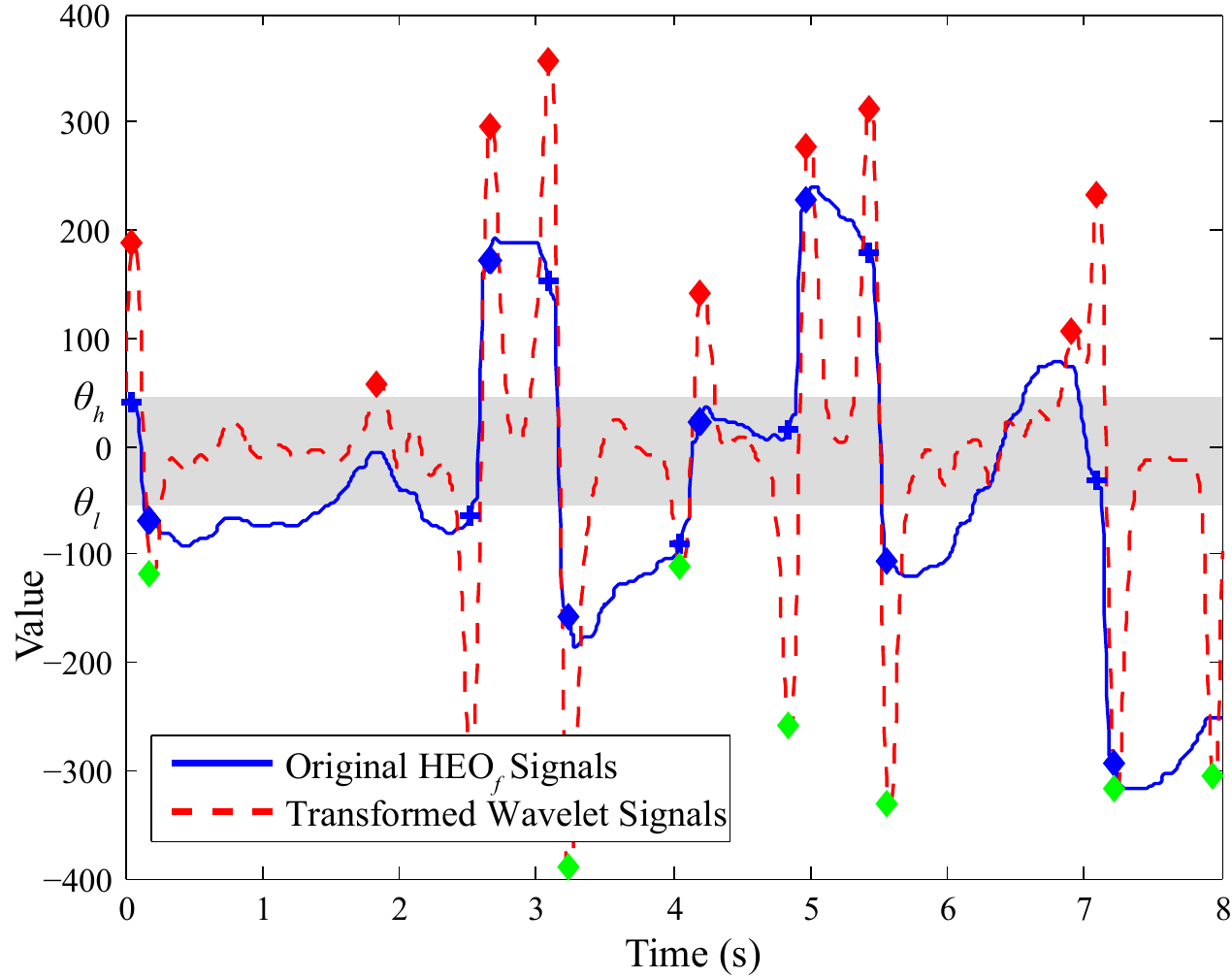}
 \caption{The saccade detected by using continuous wavelet transform. Similar to blink detection, we applied two thresholds $\theta_{h}$ and $\theta_{l}$ on the transformed wavelet signals and used peak detection on the transformed wavelet signals. Blue cross markers and diamond markers indicate the start and end points of each saccade, respectively.}
\label{fig:saccade}
\end{figure}

\subsubsection{Feature Extraction from Forehead EOG}
After preprocessing forehead EOG signals and extracting
VEO$_{\mbox{\scriptsize \emph{f}}}$ and HEO$_{\mbox{\scriptsize
    \emph{f}}}$, we detected eye movements such as blinks and saccades
using the wavelet transform method \cite{bulling2011eye}. We computed
the continuous wavelet coefficients at a scale of 8 with a Mexican hat wavelet defined by
\begin{equation}
\psi(t) = \frac{2}{\sqrt{3\sigma}\pi^{\frac{1}{4}}}(1-\frac{t^2}{\sigma^2})e^{\frac{-t^2}{2\sigma^2}},
\end{equation}
where $\sigma$ is the standard deviation. Because the wavelet
transform is sensitive to singularities, we used the peak detection
algorithm on the wavelet coefficients to detect blinks and saccades
from the forehead VEO$_{\mbox{\scriptsize \emph{f}}}$ and
HEO$_{\mbox{\scriptsize \emph{f}}}$, respectively. The detected blinks and saccades are shown in Figures \ref{fig:blink} and \ref{fig:saccade}, respectively.

By applying thresholds on the continuous wavelet coefficients, we
encoded the positive and negative peaks in forehead
VEO$_{\mbox{\scriptsize \emph{f}}}$ and HEO$_{\mbox{\scriptsize
    \emph{f}}}$ into sequences, where the positive peak was encoded as
`1' and the negative one as `0'. A saccade is characterized by a
sequence of two successive positive and negative peaks in the
coefficients. A blink contains three successive large peaks, namely,
negative, positive, and negative, and the time between two positive
peaks should be smaller than the minimum time. Therefore, for the
encoding, segments with `01' or `10' are recognized as saccade
candidates, and segments with `010' are recognized as blink
candidates. Moreover, there are some other constraints, such as slope,
correlation, and maximal segment length, for guaranteeing a precise
detection of blinks and saccades. Following the detection of blinks and saccades, we extracted the statistical parameters, such as the mean, maximum, variance, and derivative, of different eye movements with an 8 s non-overlapping window as the EOG features. We extracted a total of 36 EOG features from the detected blinks, saccades, and fixations. Table \ref{tab:table1} presents the details of the extracted 36 eye movement features.

\begin{table}[!htp]\small
\newcommand{\tabincell}[2]{\begin{tabular}{@{}#1@{}}#2\end{tabular}}
  \centering
	\begin{tabular}{ll}
		\hline\hline
		 \textbf{Group} & \textbf{Extracted Features} \\
		\hline
		\tabincell{l} {Blink}
 		& \tabincell{l} {maximum/mean/sum of blink rate\\maximum/minimum/mean of blink \\amplitude, mean/maximum of blink rate \\variance and amplitude variance \\power/mean power of blink amplitude \\blink numbers} \\
		\hline
		\tabincell{l} {Saccade}
 		& \tabincell{l} {maximum/minimum/mean of saccade \\rate and saccade amplitude, maximum/mean \\of saccade rate variance and amplitude \\variance, power/mean power of saccade \\amplitude, saccade numbers} \\
		\hline
		\tabincell{l} {Fixation}
 		& \tabincell{l} {mean/maximum of blink duration \\variance and saccade duration variance \\maximum/minimum/mean of blink \\duration and saccade duration.}\\
		\hline
		\hline
	\end{tabular}
 	\caption{The details of the extracted 36 eye movement features.}
 	\label{tab:table1}
\end{table}

\begin{figure}[!htbp]
 \centering
 \includegraphics[width=0.92\columnwidth]{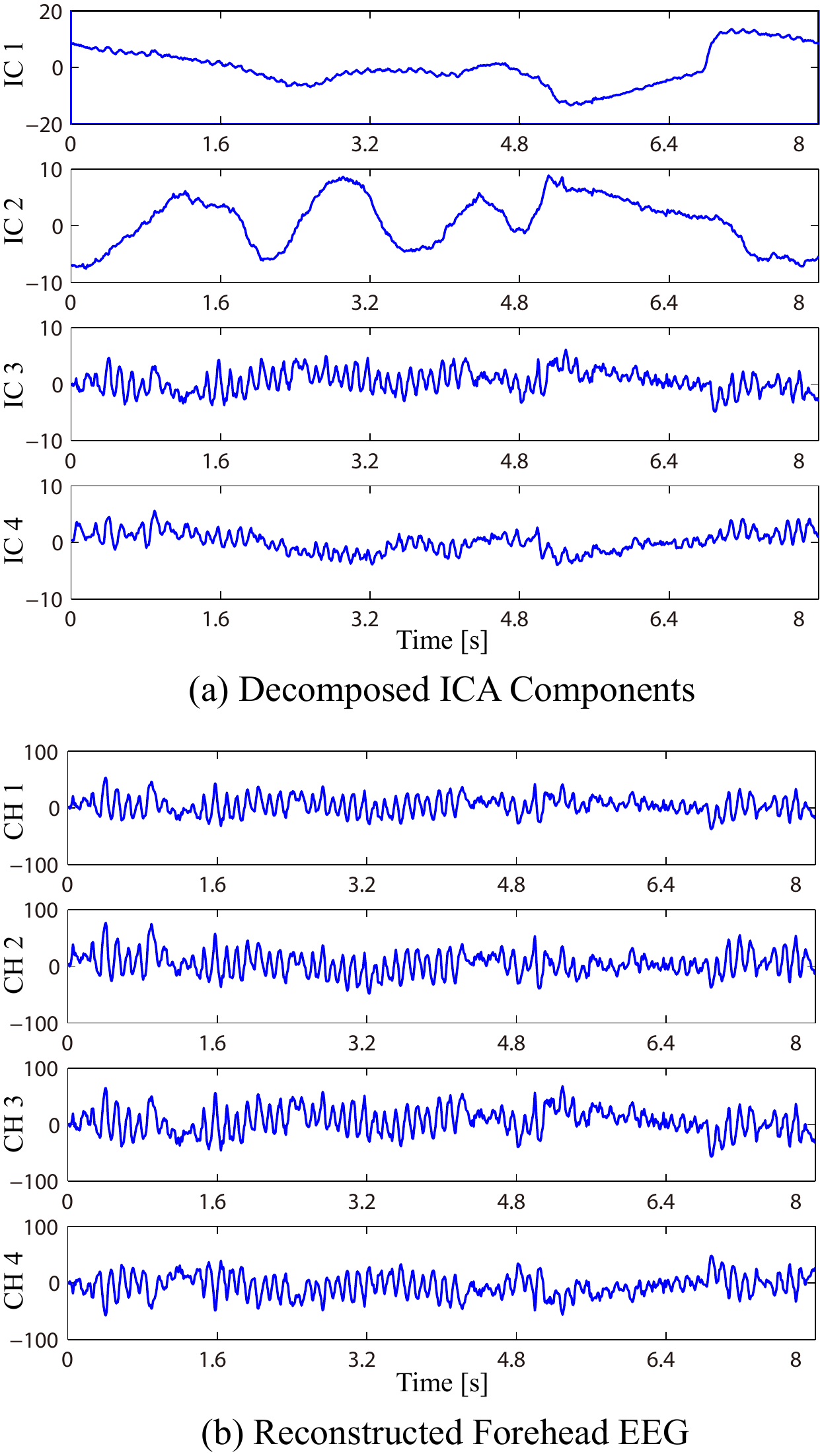}
 \caption{(a) The decomposed independent components from the four forehead channels (Nos. 4-7) using ICA. IC 1 and IC 2 are EOG components for eye activities. (b) The reconstructed forehead EEG by filtering out the EOG components. It can be observed that strong alpha activities are verified under eye closure conditions.}
\label{fig:figure3}
\end{figure}

\subsubsection{Forehead EEG Signal Extraction}
For conventional EEG-based approaches, the EOG signals are always
considered to be severe contamination, particularly for frontal
sites. Many methods have been proposed for removing eye movement and
blink artifacts from EEG recordings
\cite{uriguen2015eeg,daly2015force,delorme2004eeglab}. However, in
this study, we consider that both EEG and EOG contain discriminative
information for vigilance estimation. Our intuitive concept is that it
is possible to separate EEG and EOG signals from the shared forehead
electrodes. The main advantage of this concept is that we can leverage
the favourable properties of both EEG and EOG modalities while
simultaneously not increasing the setup cost.

We utilize the FASTICA algorithm to extract EEG and EOG components from the four forehead channels (Nos. 4-7) shown in Figure \ref{fig:figure1} (b). The ICA algorithm decomposes the multi-channel data into a sum of independent components \cite{jung2000removing}. Similar to artifact removal using blind signal separation in conventional approaches, the forehead EEG signals are reconstructed with a weight matrix by discarding the EOG components. The raw data recorded at the four forehead channels (Nos. 4-7) are concatenated as the input matrix $X$ for ICA as follows:
\begin{equation}
X=[Ch\_4; Ch\_5; -Ch\_6; Ch\_7],
\end{equation}
where the rows of the input matrix $X$ are signals $Ch\_4$, $Ch\_5$, $-Ch\_6$, and $Ch\_7$ from channels Nos. 4-7. After ICA decomposition, the un-mixing matrix $W$ can be obtained, which decomposes the multi-channel data into a sum of independent components as follows:
\begin{equation}
U=W*X,
\end{equation}
where the rows of $U$ are time courses of activations of the ICA components. The columns of the inverse matrix $W^{-1}$ indicate the projection strengths of the corresponding components. Therefore, the clean forehead EEG signals can be derived as
\begin{equation}
\widetilde{X}=W^{-1}*\widetilde{U},
\end{equation}
where $\widetilde{U}$ is the matrix of activation waveforms $U$ with rows representing EOG components set to zero.

The decomposed independent components and reconstructed forehead EEG of one segment under eye closure conditions are shown in Figure \ref{fig:figure3}. Under eye closure conditions, the alpha rhythm appears more dominant in EEG signals in previous studies \cite{papadelis2007monitoring}. From Figure \ref{fig:figure3} (a), we can observe that the first two rows are the corresponding eye movement components, and the last two rows contain EEG components with high alpha power values. The reconstructed signals contain characteristics of EEG waves, which are accompanied by high alpha bursts. The results presented in Figure \ref{fig:figure3} demonstrate the efficiency of our approach in extracting EEG signals from forehead electrodes.

\subsubsection{Feature Extraction from EEG}
In addition to forehead EOG, we recorded EEG data from temporal and posterior sites, which showed high relevance along with vigilance in the literature and our previous work \cite{khushaba2011driver,shi2013eeg}. For preprocessing, the raw EEG data were processed with a band-pass filter between 1 and 75 Hz to reduce artifacts and noise and downsampled to 200 Hz to reduce the computational complexity. For feature extraction, an efficient EEG feature called differential entropy (DE) was proposed for vigilance estimation and emotion recognition \cite{shi2013differential,duan2013differential}, which showed superior performance compared to the conventional power spectral density features.

The original formula for calculating differential entropy is defined as
\begin{equation}
h(X)=-\int_{X}f(x)log(f(x))dx.
\end{equation}
If a random variable obeys the Gaussian distribution $N(\mu,\sigma^2)$, the differential entropy can simply be calculated by the following formulation,
\begin{equation}
h(X)=-\int_{-\infty}^{\infty}f(x)log(f(x))dx=\frac{1}{2}\log{2\pi e\sigma^2},
\end{equation}
where $f(x)=\frac{1}{\sqrt{2\pi\sigma^2}}\exp{\frac{(x-\mu)^2}{2\sigma^2}}$.

According to the DE definition mentioned above, for each EEG segment, we extracted the DE features from five frequency bands: delta (1-4 Hz), theta (4-8 Hz), alpha (8-14 Hz), beta (14-31 Hz), and gamma (31-50 Hz). We also extracted the DE features from the total frequency band (1-50 Hz) with a 2 Hz frequency resolution.  All the DE features were calculated using short-term Fourier transforms with an 8 s non-overlapping window.

\subsection{Vigilance Estimation}
After obtaining vigilance labels and EOG/EEG features, we used support
vector regression (SVR) with radial basis function (RBF) kernels as a
basic regression model. The optimal values of the parameters $c$ and
$g$ were tuned with the grid search. As the modality fusion strategy,
we used feature-level fusion, in which the feature vectors of EEG and
EOG are directly concatenated into a larger feature vector as
inputs. For evaluation, we separated the entire data from one
experiment into five sessions and evaluated the performance with
5-fold cross validation. There are a total of 885 samples for each experiment.

The root mean square error (RMSE) and correlation coefficient (COR)
are the most commonly used evaluation metrics for continuous
regression models \cite{nicolaou2011continuous}. RMSE is the squared
error between the prediction and the ground truth, and it is defined as follows:
\begin{equation}
RMSE(Y,\hat{Y})=\sqrt{\frac{1}{N}\sum^N_{i=1}(y_i-\hat{y}_i)^2},
\end{equation}
where $Y=(y_1,y_2,...,y_N)^T$ is the ground truth and $\hat{Y}=(\hat{y}_1,\hat{y}_2,...,\hat{y}_N)^T$ is the prediction.

Since RMSE-based evaluation cannot provide structural information, we used COR to overcome the shortcomings of RMSE. COR provides an evaluation of the linear relationship between the prediction and the ground truth, which reflects the consistency of their trends. Pearson's correlation coefficient is defined as follows:
\begin{equation}
COR(Y,\hat{Y})=\frac{\sum^N_{i=1}(y_i-\bar{y})(\hat{y}_i-\bar{\hat{y}})}{\sqrt{\sum^N_{i=1}(y_i-\bar{y})^2\sum^N_{i=1}(\hat{y}_i-\bar{\hat{y}})^2}},
\end{equation}
where $\bar{y}$ and $\bar{\hat{y}}$ are the means of $Y$ and $\hat{Y}$. However, COR is sensitive to short segments and is appropriate for long evaluation metrics. Therefore, we concatenated the predictions and ground truth of five sessions and calculated COR as the final evaluation. In general, the more accurate the model is, the higher the COR is and the lower the RMSE is.

\subsection{Incorporating Temporal Dependency into Vigilance Estimation}
Vigilance is a dynamic changing process because the intrinsic mental
states of users involve temporal evolution. To incorporate the temporal dependency into vigilance estimation, we introduced continuous conditional neural field (CCNF) and continuous conditional random field (CCRF) when constructing vigilance estimation models. CCNF and CCRF are extensions of conditional random field (CRF) \cite{lafferty2001conditional} for continuous variable modelling that incorporates temporal or spatial information and have shown promising performance in various applications \cite{baltrusaitis2013dimensional,imbrasaite2014ccnf,baltruvsaitis2014continuous}. CCNF combines the nonlinearity of conditional neural fields \cite{peng2009conditional} and the continuous output of CCRF.

The probability distribution of CCNF for a particular sequence is defined as follows:
\begin{equation}
P(\textbf{y}|\textbf{x})=\frac{exp(\Psi)}{\int_{-\infty}^{\infty}exp(\Psi)d\textbf{y}},
\end{equation}
where $\int_{-\infty}^{\infty}exp(\Psi)d\textbf{y}$ is the normalization function, $\textbf{x}=\{x_1,x_2,\cdots,x_n\}$ is a set of input observations, $\textbf{y}=\{y_1,y_2,\cdots,y_n\}$ is a set of output variables, and $n$ is the length of the sequence.

There are two types of features defined in these models: vertex features $f_k$ and edge features $g_k$. The potential function $\Psi$ is defined as follows:
\begin{equation}
\Psi=\sum_{i}\sum_{k=1}^{K_1}\alpha_{k}f_{k}(y_i,\textbf{x}_i,\textbf{$\bm{\theta}$}_{k})+\sum_{i,j}\sum_{k=1}^{K_2}\beta_{k}g_{k}(y_i,y_j),
\end{equation}
where $\alpha_{k}>0$, $\beta_{k}>0$, the vertex features $f_k$ denote the mapping from $\textbf{x}_i$ to $y_i$ with a one-layer neural network, and $\bm{\theta}_k$ is the weight vector for the neuron $k$.

The vertex features of CCNF are defined as

\begin{equation}
f_k(y_i,\bm{x}_i,\bm{\theta}_k)=-(y_i-h(\bm{\theta}_k,\bm{x}_i))^2, and
\end{equation}

\begin{equation}
h(\bm{\theta},\bm{x}_i)=\frac{1}{1+e^{-\bm{\theta}^{T}\bm{x}_i}},
\end{equation}
where the optimal number of vertex features $K_1$ is tuned with cross-validation. In our experiments, we evaluated $K_1=\{10,20,30\}$.

The edge features $g_k$ denote the similarities between observations $y_i$ and $y_j$, which are defined as
\begin{equation}
g_k(y_i,y_j)=-\frac{1}{2}S^{(k)}_{i,j}(y_i-y_j)^2,
\end{equation}
where the similarity measure $S^{(k)}$ controls the existence of the connections between two vertices.

In the experiments, $K_2$ is set to 1 and $S^{(k)}$ is set to 1 when
two nodes $i$ and $j$ are neighbours; otherwise, it is 0. The sequence
length $n$ is set to seven. The formulas for CCRF are the same as
those for CCNF, except for the definition of vertex features. The vertex features of CCRF are defined as
\begin{equation}
f_k(y_i,\bm{x}_{i,k})=-(y_i-\bm{x}_{i,k})^2.
\end{equation}

The training of parameters in CCRF and CCNF is based on the conditional log-likelihood $P(\textbf{y}|\textbf{x})$ as a multivariate Gaussian. For more details regarding the learning and inference of CCRF and CCNF, please refer to \cite{baltruvsaitis2014continuous,imbrasaite2014ccnf}. The outputs of support vector regression are used to train CCRF, and the original multi-dimensional features are used to train CCNF. The CCRF and CCNF regularization hyper-parameters for $\alpha_{k}$ and $\beta_{k}$ are chosen based on a grid search in $10^{[0,1,2]}$ and $10^{[-3,-2,-1,0]}$ using the training set, respectively.

\section{Experimental Results}
\subsection{Forehead EOG-Based Vigilance Estimation}
First, we evaluated the similarity between forehead EOG and
traditional EOG and the performance of forehead EOG-based vigilance
estimation for different separation strategies. We extracted forehead
VEO$_{\mbox{\scriptsize \emph{f}}}$ and HEO$_{\mbox{\scriptsize
    \emph{f}}}$ using the minus and ICA separation approaches and
computed the correlation with traditional VEO and HEO. The mean
correlation coefficients of VEO$_{\mbox{\scriptsize \emph{f}}}$-MINUS,
VEO$_{\mbox{\scriptsize \emph{f}}}$-ICA, HEO$_{\mbox{\scriptsize
    \emph{f}}}$-MINUS, and HEO$_{\mbox{\scriptsize \emph{f}}}$-ICA are
0.63, 0.80, 0.81, and 0.75, respectively. These comparative results
demonstrate that the extracted forehead VEO$_{\mbox{\scriptsize
    \emph{f}}}$ and HEO$_{\mbox{\scriptsize \emph{f}}}$ contain most
of the principal information of traditional EOG.

The mean RMSE, the mean COR and their standard deviations for
different separation methods are presented in Table
\ref{tab:table2}. `ICA-MINUS' denotes ICA-based
VEO$_{\mbox{\scriptsize \emph{f}}}$ and minus-based
HEO$_{\mbox{\scriptsize \emph{f}}}$ separations, and it has the
highest correlation coefficient with traditional VEO and HEO. As shown
in Table \ref{tab:table2}, ICA-MINUS achieves the best performance for
vigilance estimation in terms of both COR and RMSE. It is consistent
with the above results that VEO$_{\mbox{\scriptsize \emph{f}}}$-ICA
and HEO$_{\mbox{\scriptsize \emph{f}}}$-MINUS are more similar to the
original VEO and HEO. For VEO, it contains many blink components, such
as impulses, which are more likely to be detected by ICA. In contrast,
the minus method reduces the amplitude of VEO signals since the polarity of the pair electrodes is the same. For HEO, saccade components are more difficult to be detected by ICA, and the polarity of the pair electrodes is different.

\begin{table}[!htbp]\small
\centering
\begin{tabular}{|c|c|c|c|c|c|}
\hline
\multicolumn{2}{|c|}{ICA-MINUS} & \multicolumn{2}{c|}{EOG-ICA} & \multicolumn{2}{c|}{EOG-MINUS} \\ \hline
COR               & RMSE              & COR           & RMSE         & COR            & RMSE          \\ \hline
\textbf{0.7773}   & \textbf{0.1188}   & 0.4774        & 0.1582       & 0.7193         & 0.1288        \\ \hline
\textbf{0.1745}   & \textbf{0.0391}   & 0.5381        & 0.0844       & 0.3492         & 0.0588        \\ \hline
\end{tabular}
\caption{The mean RMSE, the mean COR, and their standard deviations with different separation methods. Here, the numbers in the first and second rows are the averages and standard deviations, respectively.}
\label{tab:table2}
\end{table}

\subsection{EEG-Based Vigilance Estimation}

We reconstructed the frontal 4-channel EEG from the forehead signals based on the ICA algorithm. In the experiments, we also recorded 12-channel and 6-channel EEG signals from posterior and temporal sites. We extracted the DE features in two ways: one is from the five frequency bands, and the other is to use a 2 Hz frequency resolution in the entire frequency band. The mean COR, mean RMSE and their standard deviations of different EEG features from different brain areas are shown in Table \ref{tab:table3}. The ranking of the performance for EEG-based vigilance estimation from different brain areas is as follows: posterior, temporal, and forehead sites. For the single EEG modality, the posterior EEG contains the most critical information for vigilance estimation, which is consistent with previous findings \cite{khushaba2011driver,shi2013eeg}. The EEG features with a 2 Hz frequency resolution achieve better performance than those with five frequency bands. In the later experimental evaluation in this paper, we employ the EEG features with a 2 Hz frequency resolution of the entire frequency band.

\begin{table}[!b]
 \centering
  \includegraphics[width=0.95\columnwidth]{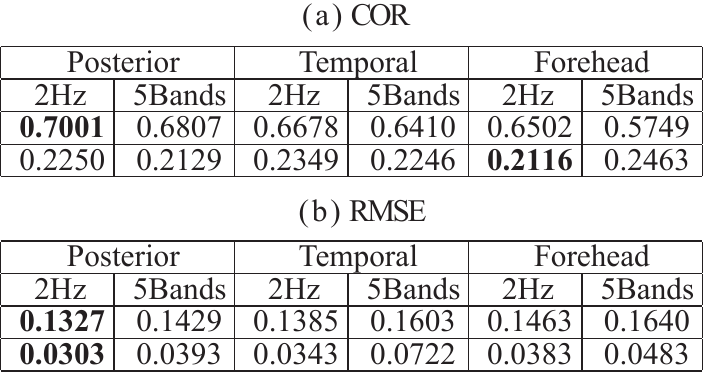}
  \caption{The average and standard deviations of COR and RMSE for different EEG features. Here, the numbers in the first and second rows are the averages and standard deviations, respectively.}
  \label{tab:table3}
\end{table}

In addition to the accuracy that we discussed above for decoding brain
states, another important concern is to examine whether patterns of
brain activity under different cognitive states exist and whether
these patterns are to some extent common across
individuals. Identifying the specific relationship between brain
activities and cognitive states provides evidence and support for
understanding the information processing mechanism of the brain and
brain state decoding \cite{haynes2006decoding}. Huang \textit{et al.}
demonstrated the specific links between changes in EEG spectral power
and reaction time during sustained-attention tasks
\cite{huang2009tonic}. They found that significant tonic power
increases occurred in the alpha band in the occipital and parietal
areas as reaction time increased. Ray and colleagues proposed that
alpha activities of EEG reflect attentional demands and that beta activities reflect emotional and cognitive processes \cite{ray1985eeg}. They found increasing parietal alpha activities for tasks that do not require attention.

\begin{figure}[!htbp]
 \centering
 \includegraphics[width=\columnwidth]{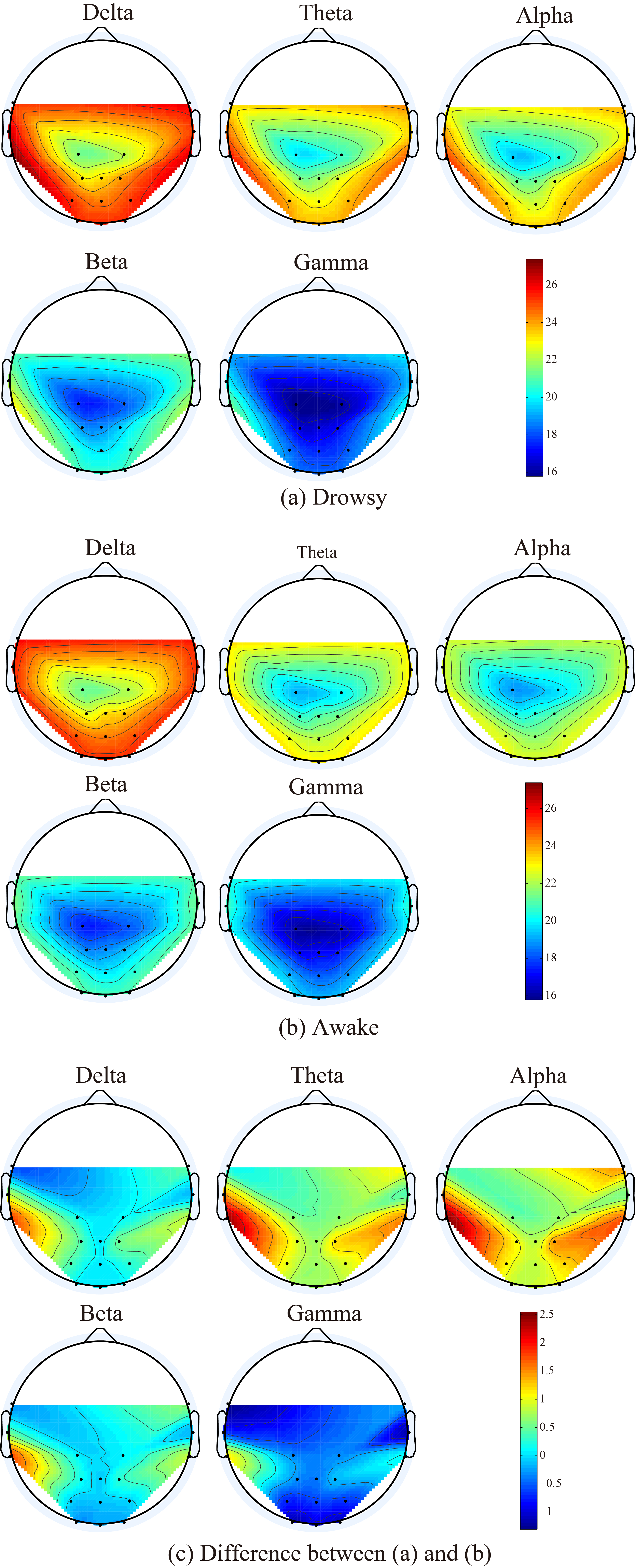}
 \caption{The mean neural patterns of awake and drowsy states as well as the difference between these two states. By applying two thresholds (0.35 and 0.7) to the PERCLOS index, we split the EEG data into three categories: awake, tired, and drowsy. From the average neural patterns, we observe that drowsy states have higher alpha frequency activities in parietal areas and lower gamma frequency activities in temporal areas.}
\label{fig:patterns}
\end{figure}

In this work, to investigate the changes in neural patterns associated
with vigilance, we split the EEG data into three categories (awake,
tired, and drowsy) with two thresholds (0.35 and 0.7) according to the
PERCLOS index. We averaged the DE features over different
experiments. Figure \ref{fig:patterns} presents the mean neural
patterns of awake and drowsy states as well as the difference between
them. As shown in Figure \ref{fig:patterns},  increasing theta and
alpha frequency activities exist in parietal areas and decreasing
gamma frequency activities exist in temporal areas in drowsy states in
contrast to awake states. These results are consistent with previous
findings in the literature
\cite{ray1985eeg,davidson2007eeg,huang2009tonic,o2009uncovering,peiris2011detection,lin2013can,martel2014eeg}
and support the previous evidence that the increasing trend for the ratio of slow and fast waves of EEG activities reflects decreasing attentional demands \cite{jap2009using}.

%

\subsection{Modality Fusion with Temporal Dependency}

\begin{figure}[!b]
 \centering
 \includegraphics[width=\columnwidth]{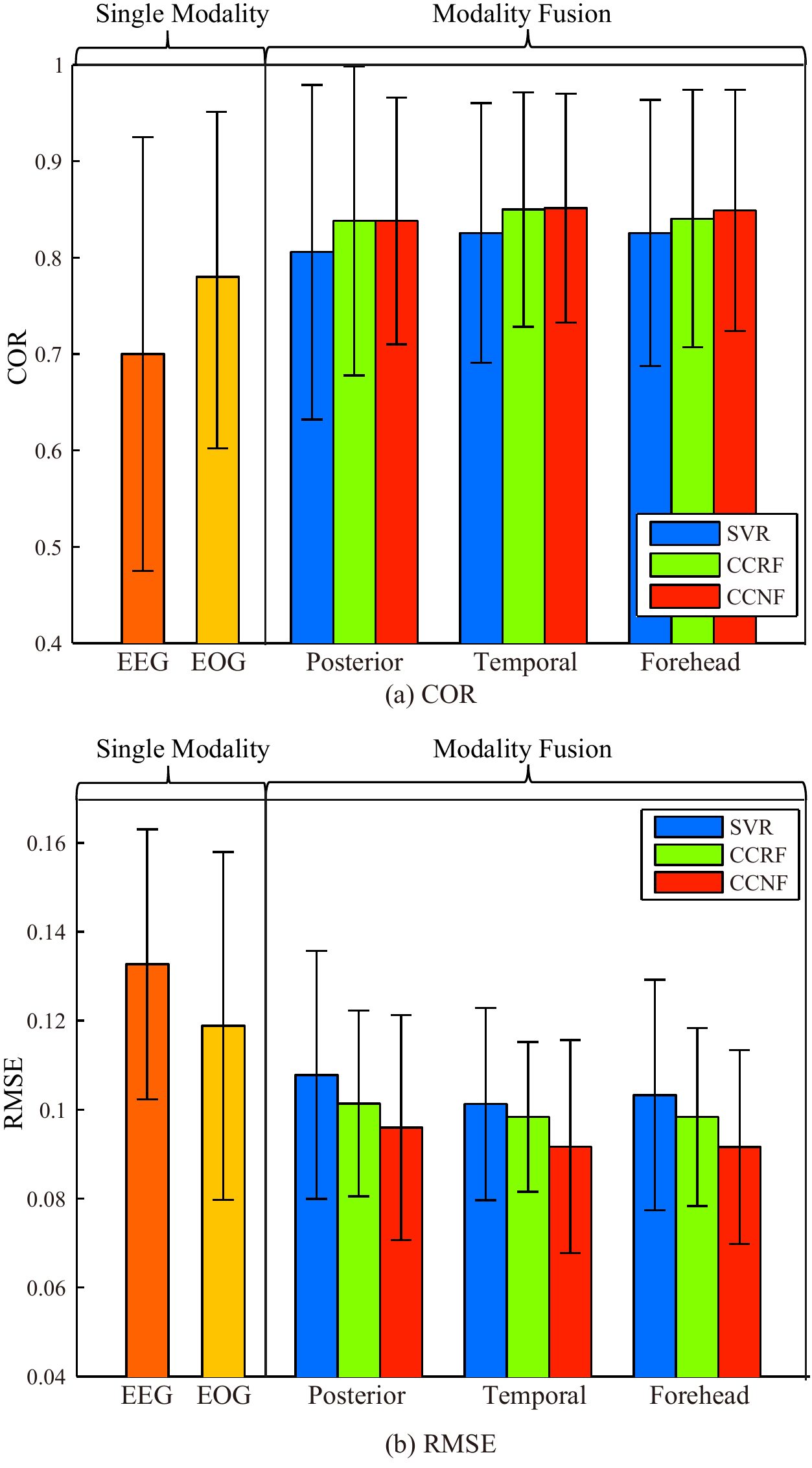}
 \caption{The mean COR and mean RMSE of each single modality and different modality fusion strategies.}
\label{fig:figure5}
\end{figure}

In this section, we introduced a multimodal vigilance estimation
approach with the fusion of EEG and forehead EOG. We combined the EEG
signals from different sites (forehead, temporal, and posterior) and
forehead EOG signals to utilize their complementary characteristics
for vigilance estimation. The performance of each single modality and
different modality fusion strategies are shown in Figure
\ref{fig:figure5}. For a single modality, forehead EOG achieves better
performance than posterior EEG. The reason for this result is that
forehead EOG has more information in common with the annotations of
eye tracking data. Modality fusion can significantly enhance the
regression performance in comparison with a single modality with a
higher COR and lower RMSE. We evaluated the statistical significance
using one-way analysis of variance (ANOVA), and the $p$ values of COR for forehead EOG and posterior EEG are 0.2978 and 0.0264, respectively. The $p$ values of RMSE for forehead EOG and posterior EEG are 0.0654, and 0.0002, respectively.

For different brain areas, an interesting observation is that the
fusion of forehead EOG and forehead EEG achieves better performance
than that of forehead EOG and posterior EEG, whereas for single EEG,
the posterior site achieves the best performance. These results
indicate that forehead EEG and forehead EOG have more coherent
information. The temporal EEG performs slightly better than the
forehead EEG. However, the former requires six extra electrodes for
the setup. The forehead setup only uses four shared electrodes and
both EOG and EEG features can be extracted. Therefore, the information
flow can be increased without any additional setup cost. From the
above discussion, we see that the forehead approach is preferred for real-world applications.

\begin{figure*}[!htbp]
 \centering
 \includegraphics[width=1.64\columnwidth]{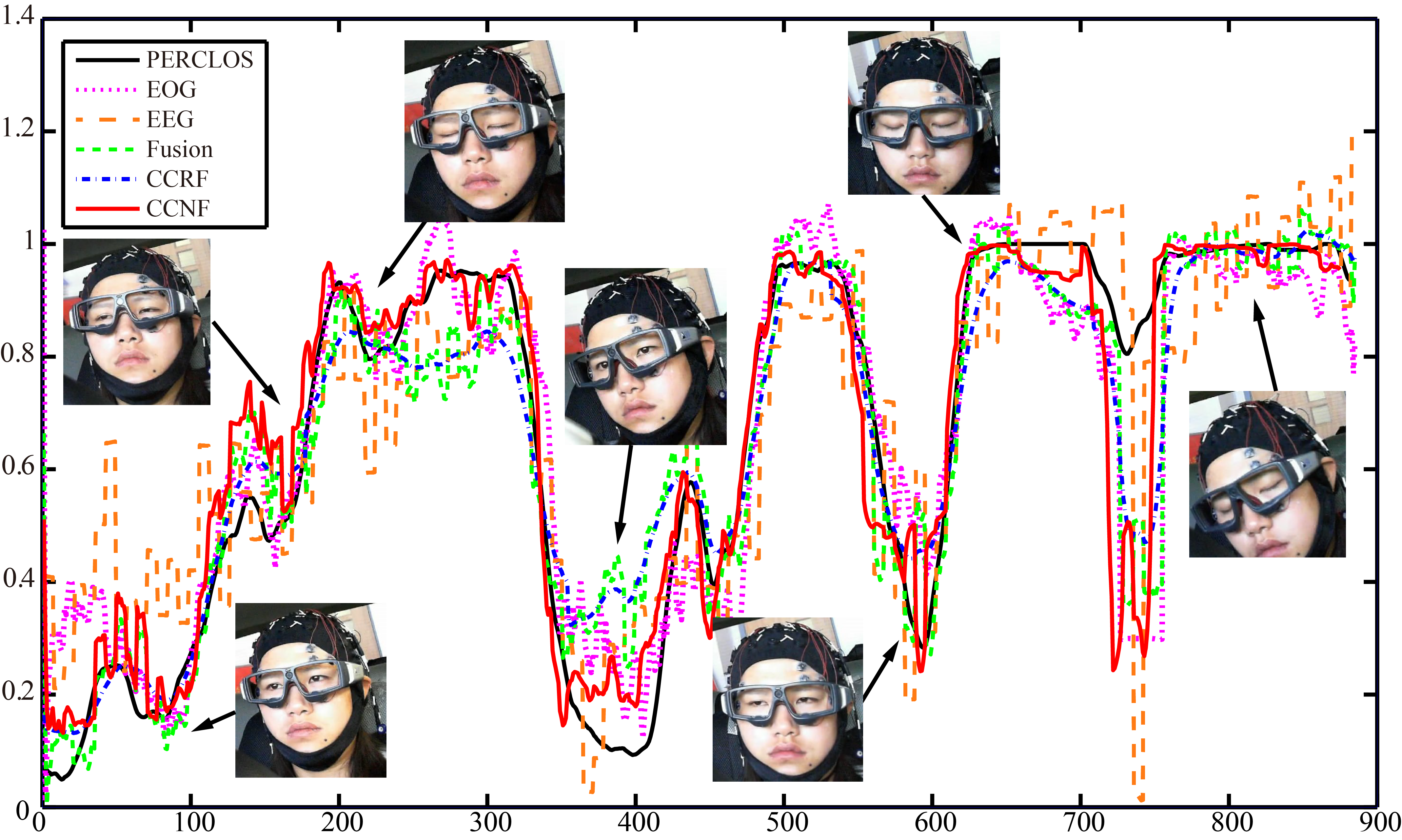}
 \caption{The continuous vigilance estimation of different methods in
   one experiment. As shown, the predictions from our proposed approaches are almost consistent with the true subjects' behaviours and cognitive states.}
\label{fig:figure7}
\end{figure*}

To incorporate temporal dependency information into vigilance
estimation, we adopted CCRF and CCNF in this study. As shown in Figures \ref{fig:figure5} (a) and (b), the temporal dependency models can enhance the performance. For the forehead setup, the mean COR/RMSE of SVR, CCRF, and CCNF are 0.83/0.10, 0.84/0.10, and 0.85/0.09, respectively. The CCNF achieves the best performance with higher accuracies and lower standard deviations.

To verify whether the predictions from our proposed approaches are consistent with the true subjects' behaviours and cognitive states, the continuous vigilance estimation of one experiment is shown in Figure \ref{fig:figure7}. The snapshots in Figure \ref{fig:figure7} show the frames corresponding to different vigilance levels. We can observe that our proposed multimodal approach with temporal dependency can moderately predict the continuous vigilance levels and its trends.

%

\begin{figure}[!t]
 \centering
 \includegraphics[width=\columnwidth]{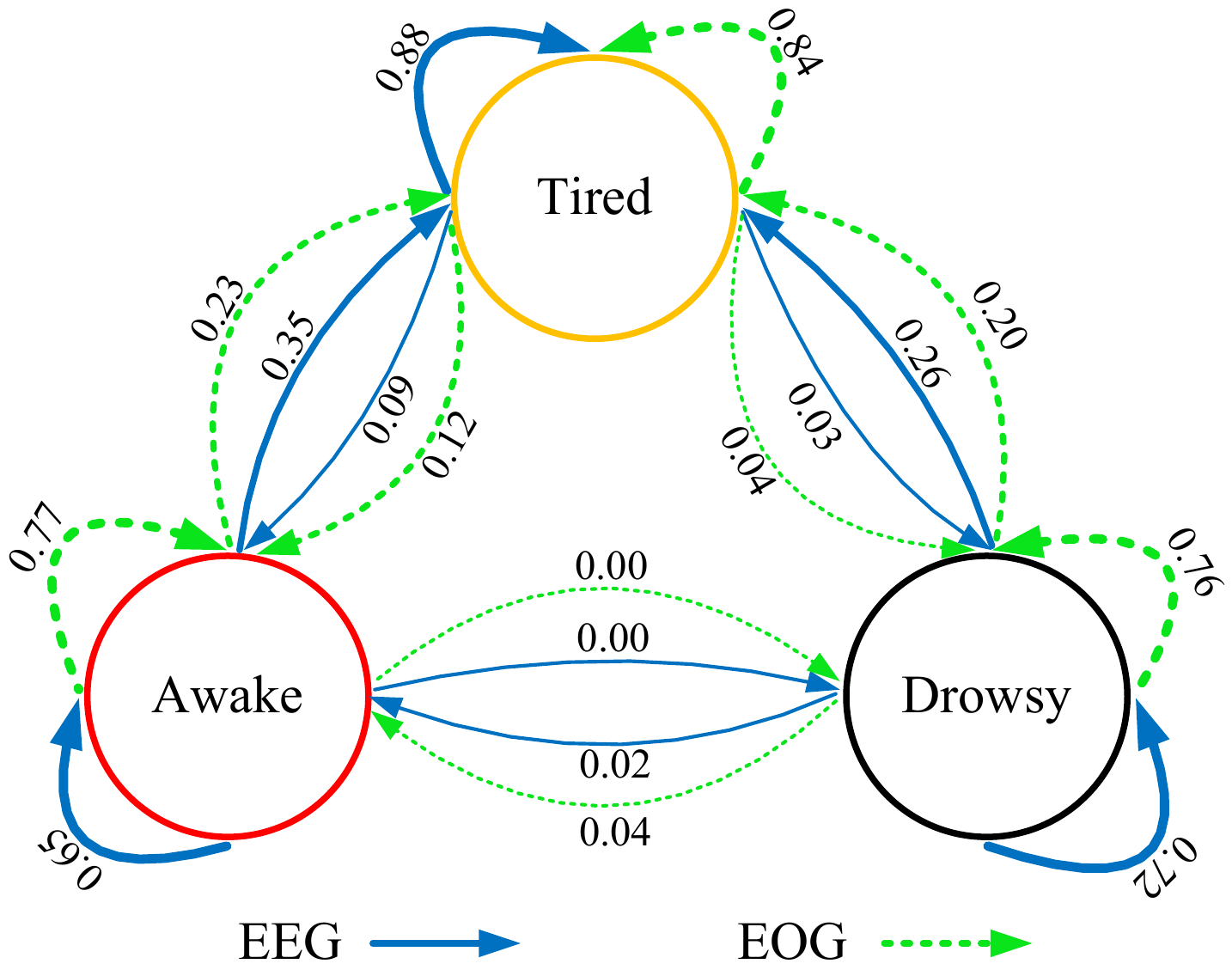}
 \caption{Confusion graph of forehead EOG and posterior EEG, which shows their complementary characteristics for vigilance estimation. Here, the numbers denote the percentage values of samples in the class (arrow tail) classified as the class (arrow head). Bolder lines indicate higher values.}
\label{fig:ConfusionPicture}
\end{figure}

To further investigate the complementary characteristics of EEG and
EOG, we analysed the confusion matrices of each modality, which
reveals the strength and weakness of each modality. We split the EEG
data into three categories, namely, awake, tired and drowsy
states, with thresholds according to the corresponding PERCLOS index
as described above. Figure \ref{fig:ConfusionPicture} presents the
mean confusion graph of forehead EOG and posterior EEG of all
experiments. These results demonstrate that posterior EEG and forehead
EOG have important complementary characteristics. Forehead EOG has the
advantage of classifying awake and drowsy states (77\%/76\%) compared
to the posterior EEG (65\%/72\%), whereas posterior EEG outperforms
forehead EOG in recognizing tired states (88\% vs. 84\%). The forehead
EOG modality achieves better performance than the posterior EEG
overall. This result may be because our ground truth labels are obtained with eye movement parameters from eye tracking glasses. The forehead EOG contains more similar information with the experimental observations. Moreover, awake states and tired states are often misclassified with each other, and similar results are observed for drowsy and tired states. In contrast, awake states are seldom misclassified as drowsy states and vice versa for both modalities. These observations are consistent with our intuitive knowledge. EEG and EOG features of awake and drowsy states should have larger differences. These results indicate that EEG and EOG have different discriminative powers for vigilance estimation. Combining the complementary information of these two modalities, modality fusion can improve the prediction performance.

\section{Discussion}
In this study, we have developed a multimodal approach for vigilance
estimation regarding temporal dependency and combining EEG and
forehead EOG in a simulated driving environment. Several researchers
have performed pilot studies for on-road real driving tests. Papadelis
\textit{et al.} designed an on-board system to assess a driver's
alertness level in real driving conditions
\cite{papadelis2007monitoring}. They found that EEG and EOG are
promising neurophysiological indicators for monitoring
sleepiness. Haufe \textit{et al.} performed a study to assess the
real-world feasibility of EEG-based detection of emergency braking
intention \cite{haufe2014electrophysiology}. Indeed, in addition to
driving applications, there are many other scenarios that require
vigilance estimation, such as students' performance in classes. Hans and colleagues examined how cognitive fatigue influences students' performance on standardized tests in their study \cite{sievertsen2016cognitive}. To evaluate the feasibility of our approach, we will apply our vigilance estimation approach to real scenarios in future work.

Considering the wearability and feasibility of a vigilance estimation
device for real-world applications, we have designed four-electrode
placements on the forehead, which are suitable for attachment in a
wearable headset or headband. We can collect both EEG and EOG
simultaneously and combine their advantages via shared forehead
electrodes. The experimental results demonstrate that our proposed
approach can achieve comparable performance with the conventional
methods on critical brain areas, such as parietal and occipital
sites. This approach increases the information flow with easy setups
while not considerably increasing the cost.

In recent years, substantial progress has been made in dry electrodes
and high-performance amplifiers. Several commercial EEG systems have
emerged for increasing the usability in real-world applications \cite{grozea2011bristle,hairston2014usability,mullen2015real}. It is feasible to integrate these techniques with our proposed approach to design a new wearable hybrid EEG and forehead EOG system for vigilance estimation in the future.

In this study, we focus only on vigilance estimation without
considering any neurofeedback. For example, a feedback can be timely
provided to the driver to enhance driving safety if the vigilance
detection system indicates that he or she is in an extremely tired
state. An adaptive closed-loop BCI system that consists of vigilance detection and feedback is very useful in changing environments \cite{wu2010optimal,lin2013can}. How to efficiently provide and assess the feedback in high vigilance tasks should be further investigated.

Due to individual differences of neurophysiological signals across subjects and sessions, the performance of vigilance estimation models may be dramatically degraded. The generalization performance of vigilance estimation models should be considered for individual differences and adaptability. However, training subject-specific models requires time-consuming calibrations. To address these problems, one efficient approach is to train models on the existing labelled data from a group of subjects and generalize the models to the new subjects with transfer learning techniques \cite{pan2010survey,wronkiewicz2015leveraging,morioka2015learning,zhang2015transfer,zheng2016personalizing}.

\section{Conclusion}
In this paper, we have proposed a multimodal vigilance estimation
approach using EEG and forehead EOG. We have applied different
separation strategies to extract VEO$_{\mbox{\scriptsize \emph{f}}}$,
HEO$_{\mbox{\scriptsize \emph{f}}}$ and EEG signals from four shared
forehead electrodes. The COR and RMSE of single forehead EOG-based and
EEG-based methods are 0.78/0.12 and 0.70/0.13, respectively, whereas
the modality fusion with temporal dependency can significantly enhance
the performance with values of 0.85/0.09. The experimental results
have demonstrated the feasibility and efficiency of our proposed
approach based on the forehead setup. Our vigilance estimation method
has the following three main advantages: both EEG and EOG signals can
be acquired simultaneously with four shared electrodes on the
forehead; modelling both internal cognitive states and external
subconscious behaviours with fusion of forehead EEG and EOG; and
introducing temporal dependency to capture the dynamic patterns of the
vigilance of users. From the experimental results, we have observed
that phenomena of increasing theta and alpha frequency activities and
decreasing gamma frequency activities in drowsy states do exist in contrast to awake states. We have also investigated the complementary characteristics of forehead EOG and EEG for vigilance estimation. Our experimental results indicate that the proposed approach can be used to implement a wearable passive brain-computer interface for tasks that require sustained attention.

\section*{Acknowledgments}
This work was supported in part by grants from the National Natural
Science Foundation of China (Grant No. 61272248), the National Basic
Research Program of China (Grant No. 2013CB329401), and the Major Basic Research Program of Shanghai Science and Technology Committee (15JC1400103).

\section*{References}
\bibliographystyle{dcu}
\bibliography{vigilance}

\end{document}